# A Linguistically Driven Framework for Query Expansion via Grammatical Constituent Highlighting and Role-Based Concept Weighting

B. Selvaretnam
Faculty of IT
Monash University

M. Belkhatir
Faculty of Computer Science
University of Lyon
mohammed.belkhatir@univ-lyon1.fr

## Abstract

In this paper, we propose a linguistically-motivated query expansion framework that recognizes and encodes significant query constituents that characterize query intent in order to improve retrieval performance. Concepts-of-Interest are recognized as the core concepts that represent the gist of the search goal whilst the remaining query constituents which serve to specify the search goal and complete the query structure are classified as descriptive, relational or structural. Acknowledging the need to form semantically-associated base pairs for the purpose of extracting related potential expansion concepts, an algorithm which capitalizes on syntactical dependencies to capture relationships between adjacent and non-adjacent query concepts is proposed. Lastly, a robust weighting scheme that duly emphasizes the importance of query constituents based on their linguistic role within the expanded query is presented. We demonstrate improvements in retrieval effectiveness in terms of increased mean average precision (MAP) garnered by the proposed linguistic-based query expansion framework through experimentation on the TREC ad hoc test collections.

## 1 Introduction

Natural language is not always well-structured and may be semantically ambiguous, thus making it difficult to formulate the most appropriate and restrictive search query that is in line with the vocabulary of the documents being searched for. In recognition of this problem, several query expansion efforts have emerged over the years in an attempt to minimize query-document vocabulary mismatch. However, the query expansion process requires the comprehension of the intended search goal through the identification of key concepts prior to spawning and integrating additional terms to an initial query.

It is, however, inherent that query constituents possess two distinct functionalities which are disregarded in the aforementioned research efforts. Concepts either represent the content in accordance to the search goal or serve to connect query constituents. This is shown for example in the query "*coping* (verb) *with* (preposition) *overcrowded* (verb) *prisons* (noun)" where the noun *prisons* provides the content whilst the verb *coping* and adjective *overcrowded* give specificity to the search goal. It is therefore necessary to acknowledge that query constituents take on multiple roles which if recognized and encoded appropriately would render a more accurate understanding of the intended search goal. In particular, the recognition of formal non-compositional phrases (i.e. phrasal verbs, modals, fixed phrases, idioms, collocations, proper names and acronyms) which exist within queries is of particular importance.

Furthermore, the very nature of natural language dictates that there are intrinsic relationships between adjacent and non-adjacent concepts that reveal semantic notions pertaining to a search goal. However, earlier works fail to fully capitalize on these relationships between query terms which if considered appropriately would improve retrieval performance (Song et al, 2008). Term dependencies are formalized in (Metzler & Croft, 2005) and applied in query expansion in (Metzler & Croft, 2007). In (Metzler & Croft, 2007), it is reported that retrieval performance through sequential modeling improves significantly in comparison to unigram language model (full independence) and some improved effectiveness is seen in about 65-80% of the queries examined. We postulate that retrieval per-formance is possibly hindered by the fact that only relationships between adjacent concepts are considered. One might argue that both adjacent and non-adjacent dependencies can be captured through full dependence modeling. Contrarily, this will prove costly, in especially long queries, as multiple concept pairs will be derived, from which possibly a large number would not be very meaningful. Query expansion based on these pairs would generate unrelated concepts which in turn cause digression from the original search goal. Another imperative fact is also that the dependence models highlighted in (Metzler & Croft, 2005) are purely concerned with position and proximity of concepts. We hypothesize that both adjacent and non-adjacent association among query concepts can be effectively capitalized from syntactical dependencies within queries. This then translates into meaningful query concept pairing for expansion. We exemplify this conviction through the following discussion of the queries which consist of multiple concepts representing a single information need (e.g. US President Barack Obama's inaugural address) and multiple concepts representing multiple information needs (e.g. "Barack Obama's policies and inaugural address"). Apart from recognizing that the terms *Barack* and *Obama* form a concept representing a specific person; it is intuitively evident that *US* and *President* refer to *Barack Obama*, in the former, whilst *policies* and *inaugural address* are linked to *Barack Obama*, in the latter.

Finally, a crucial element in query expansion is the process of weighting the original and expansion concepts to adequately reflect the search goal of a query. The choice of concept weighting scheme for query expansion models in existing approaches are very much dependent on the retrieval model utilized. The drawback of such models is that key concepts are established based on the statistical occurrence of a concept which is not necessarily reflective of the search goal of a query. Instead, we believe concepts should be given due emphasis based on the role it plays within a query in representing the information need

**Our contributions**: Our goal in this paper is to improve retrieval performance through query expansion by capitalizing on linguistic characteristics of queries. Queries are composed of multiple concepts of varying parts-of-speech and grammatical relations. These syntactical characteristics play distinct roles, whether it be representing the query content or providing links between the concepts, thus emphasizing semantic attributes of an expressed information need. To this end, we present a framework that consists of three major elements. We first propose a linguistically-motivated scheme for recognizing and encoding significant query constituents that characterize the intent of a query. We then extract potential expansion concepts through a proximity-based statistical query expansion method that capitalizes on grammatically linked base pairs of query concepts. Finally, we reconcile original and expansion concepts through a robust weighting scheme that is reflective of the role types of query constituents in representing an information need. One important factor is that, contrarily to state-of-the-art solutions, we are not dependent on the processed datasets to generate expansion concepts but instead consider a non-domain specific knowledge base (i.e. an n-gram corpus).

In Section 2, we cover the related works then discuss in Section 3 the details of the proposed framework, whilst in Section 4, we explain the experimental setup of the retrieval experiments, as well as the results obtained from the evaluation of our proposed scheme.

## 2 Related Work

Depending on their familiarity with the search process, users construct queries which are both short and straight to the point or long-winded (Aula, 2003). These queries may take the form of grammatically correct sentences or merely a group of keywords associated to their search goals. This basic variance of query formats motivates the need to analyze the structure and varying lengths of queries in order to decipher the intended search goal. However, as far as query structure is concerned, even though linguistic characteristics such as parts-of-speech were utilized for the purpose of key concept identification (Cao et al., 2005), it was not considered when extracting expansion concepts.

Voorhees (1994) highlights that long queries, if not well detailed, would also benefit from query expansion as much as short queries. This is similar to Lau & Goh (2006) who theorize that as query length increases, there is a higher probability that users would encounter unsuccessful searches. Di Buccio et al (2014) performed query expansion on all query terms upon distinguishing short and verbose queries with promising results. However, long queries may not always have done well with direct query expansion as such queries may consist of multiple important concepts. A verbose query may not always contain explicit information in the description itself to indicate which of these concepts is more important. Secondly, such queries may contain two or more equally essential key concepts. It is thus vital to identify the key concept expressed in a query to ensure the main search goal is emphasized (Bendersky & Croft, 2008).

Recognizing that linguistic processing may be error-prone due to semantic ambiguities that occur in natural language queries, current research (e.g. Bendersky et al. 2009; Huston and Croft 2010, Huston and Croft 2014) incorporate a combination of syntactical analysis and statistical mining in key concept identification tasks. In several existing query expansion frameworks, nouns are assumed to be of significance [(Voorhees, 1994), (Kim et al.,2004), (Cao et al., 2005), (Liu et al., 2005), (Xu et al, 2006)], while more often all non-stop words query terms were assumed to be representative of query content [(Voorhees,1994), (Abdelali et al., 2007), (Jing et al., 2005), (Gao et al. 2004), (Lioma & Ounis, 2008), (Metzler & Croft, 2007), (Song et al,  2008), (Bhogal et al., 2007), (Liu et al., 2004)], similar to that of unigram models in early information retrieval efforts. Several other works on key concept extraction such as in documents [(Eibe et al, 1999), (Turney, 2000), (Grineva et al.,2009)] and queries [(Lioma & Ounis, 2008), (Bendersky & Croft, 2008)] also emphasize on noun phrases and key concepts are established through the examination of frequency of occurrence of terms in documents, ngrams, query logs etc. (Hasan & Ng, 2014).

Thus far, concept weighting when performed independent of or alongside the retrieval model are purely statistical and have placed most emphasis on the frequency of concepts within a document corpus either through simplistic term and document frequency computation (such as in (Song et al, 2008) or through supervised learning mechanisms with multiple features but are also centered on the frequency of the concept within a variety of sources such as n-grams and query logs [(Bendersky & Croft, 2008), (Zhao et. al, 2014), (Paik & Oard, 2014)].

## 3    Outline of the Framework

In this section, we present our query expansion framework i.e. Linguistic-Statistical Query Expansion Framework (LSQE). Figure 1 depicts an overview of the stages of processing a query will be subjected to within the framework. The first step of processing is Query Segmentation which entails two steps: Non-Compositional Phrase (NCP) detection (Section 3.1.1) and Concept-Role Mapping (Section 3.1.2). The NCP detection process involves the identification and isolation of special words (e.g. proper nouns) and groups of words which must co-exist to carry a certain meaning (e.g. phrases such as idioms, phrasal verbs etc). The concept-role mapping process includes the task of identifying grammatical dependencies between concepts in a query and assigning specific role-types to each concept based on its individual functionality. The next step of processing is Extraction where grammatically linked base pairs of concepts are then derived for the purpose of extracting and selecting potential expansion concepts (Section 3.2). Lastly, a role-based concept weighting scheme is employed to assign weights to the expanded query consisting of both original query terms and relevant expansion terms (Section 3.3). The detailed steps of processing the example query "*coping with overcrowded prisons*" will be covered in the respective sections.

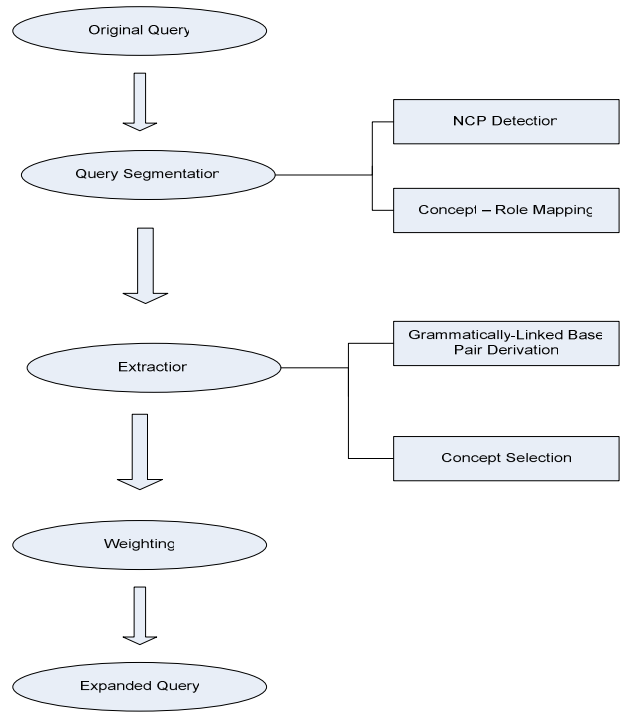

**Figure 1** General organization of the framework and processing flow of an example query

## 3.1  Query Segmentation

Query segmentation consists of (i) Non-Compositional Phrase (NcP) Detection and (ii) Concept-Role Mapping.

### 3.1.1 Non-Compositional Phrase Detection (NcP)

NcPs need to be treated differently than single word concepts within a query as their meaning is not obvious from the sense of individual terms. Several different types of non-compositional phrases exist within the English Language e.g. *phrasal verbs* (e.g. step down, take off), *modals* (e.g. be about to, had better), *fixed phrases* (e.g. in the light of), *idioms* (e.g. storm in a teacup, a bird in the hand is worth two in the bush), *collocations* (e.g. fountain pen, crystal clear), *discourse markers* (e.g. be that as it may), *proper names* (e.g. Los Angeles, George Bush) and *acronyms* (e.g. NEP - New Economic Policy; UN - United Nations). Among these, it is essential to isolate and process five non–compositional phrases i.e. *phrasal verbs, idioms, collocations, proper names* and *acronyms/abbreviations* as they directly contribute towards the search goal of the query unlike modals, fixed phrases and discourse markers which serve to complete a sentence structure and are typically captured by most linguistic parsers [(Morenberg, 2001), (Covington, 2001), (Meyer, 2009)]. The NcPs within a query are isolated (capitalization of the initial letter of each component) by comparing against a knowledge source (i.e. Wordnet ontology) as well as compiled NCP listings from the web. With NcP detection, the proper name "*United States*" would be isolated as a phrase in the query "*demographic shifts in the United States*" whilst the query "*coping with overcrowded prisons*" does not contain any non-compositional phrase. Case-folding is performed to convert all non-NCPs into lower case. This is to avoid misinterpretations by linguistic parsers which tend to assume a term/phrase is an NCP simply because it is capitalized. The algorithm summarizing the NCP detection process is shown in Algorithm 3.1.1.

**Algorithm 3.1.1** Algorithm for NCP detection

```
 Requires: Q, the set of queries
1: FOR all q∈ Q do
2:       strSegment = splitQuery(q); // strSegment is a vector of query components
3:       FOR each element strSegment[i]
4:             IF strSegment[i] is found in the NCP bank
5:                   Capitalize strSegment[i] in q
6:             ELSE
7:                   Case fold strSegment[i] in q
8:             END IF
9:       END FOR
10: END FOR
```

NCPs are detected by comparing against an ontology as well as compiled NCP lists from the web. Isolation of NCPs is achieved through capitalization and case-folding of non-NCPs.

### 3.1.2 Concept-Role Mapping

A query consists of constituents that play the role of either content or function words. Content words (e.g. nouns, verbs, adjectives and adverbs) refer to actions, objects and ideas in the everyday world. Function words (e.g. determiners, auxiliaries pronouns, conjunctions, prepositions), on the other hand, express grammatical features like mood, definiteness and pronoun reference or facilitate grammatical processes like rearrangement, compounding and embedding (Morenberg, 2001). Typically, nouns represent the subject of a sentence through names of persons, places, things and ideas whilst verbs are actions which are representative of a predicate. Other constituents such as determiners and adjectives modify nouns whilst adverbs generally alter verbs. Similarly, prepositions take on a modifying role, acting as adjectives/adverbs, by describing a relationship between words. Conjunctions, on the other hand, connect parts of a sentence or specifically the content words.

However, this crude distinction of content and function words based on parts-of-speech renders all nouns, verbs, adjectives and adverbs as key concepts. However, closer scrutiny of queries reveal that content words (e.g. nouns, adjectives) sometimes serve as modifiers and complements which complete the meaning of a query, thus may not necessarily be key concepts.

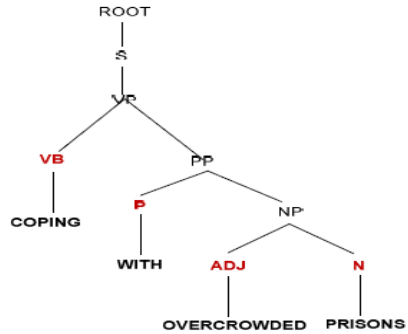

**Figure 2** Parser Output for Query "*Coping with overcrowded prisons*"

To illustrate, we refer to Fig. 2 which displays the grammatical structure of the example query "*coping with overcrowded prisons*". The parts-of-speech of the concepts are identified where "*coping*" is a verb (VB), "*with*" is a preposition (P), "*overcrowded*" is an adjective (ADJ) and "*prisons*", a noun (N). The content words in this case would be "*coping*", "*overcrowded*" and "*prisons*" whilst "*with*" takes the role of a function word. However, it would be more precise to say that "*prisons*" is the key concept of this query; "*overcrowded*", an adjectival modifier, describes the state of the prisons whilst the word "*with*" connects both words to reflect the act of dealing with prison conditions. The existence of modifiers and complements indicates that there is a need to further specify the role of the various parts-of-speech to more accurately capture the intent of a given query for more effective query expansion. It is our postulation that query constituents may be categorized into four types of concepts (which we refer to as *role-type* from this point on) that characterize the role of the constituents within a query:

Concepts can be classified within four types:

    i. **Concept-of-Interest (CoI)** which are the key concepts in a query indicative of the search goal.

    ii. **Descriptive Concept (Dc)** which describes the key concepts in further detail.

    iii. **Relational Concept (Rc)** which provides the link between concepts in a query.

    iv. **Structural Concept (Sc)** which is a stop word that helps form the structure of a query.

A simplistic approach of role type categorization based only on parts-of-speech would result in the mislabeling of role type. Both adjacent and non-adjacent pairs of concepts should be considered in the concept-role mapping process. Thus, the need for the utilization of dependency grammars (Covington, 2001; Nivre, 2005) to discover pair-wise dependencies between individual query constituents becomes evident. Pair-wise dependencies highlight the connectivity between individual constituents in which the dependent is the modifier, object or complement. Grammatical relations or typed dependencies are extracted by finding the "head" and "dependent" of constituent phrases within a query. A "head" is a word upon which everything in a phrase is centered whilst a "dependent" is every other word that is associated to the "head" (Meyer, 2009). Stop words are retained during this process to correctly identify the role of concepts by preserving the original structure of the query. Appropriate selection of "heads" in a query would lead to semantically associated "dependents" which when utilized in conjunction with the isolation of NcPs can be used to determine the appropriate assignment of CoI, Dc, Rc and Sc.

Most existing parsers provide phrase structure parse output, i.e. the nesting of multiword constituents such as noun phrase, verb phrase etc. However, syntactic parsing is not only crucial for the discovery of constituent representations but also of grammatical relations within a natural language sentence or query [(Morenberg, 2001), (McCarthy & Carroll, 2003), (Punyakanok et al., 2008)]. In recent years, grammatical relations are being included to extract typed dependencies as it is seen to be beneficial for NLP tasks (Marneffe et al., 2006).

Typed dependencies essentially consist of three main grammatical relation groupings: auxiliary, argument and modifier. The auxiliary dependencies consist of relations between words which are connected by stop words (e.g. conjunctions, determiners etc). The argument category of dependencies draws attention to subjects, objects, and complements within a sentence. The modifier dependencies highlight several parts-of-speech based modifiers (e.g. adjectival, noun compound etc) which describe and complete the meaning of other constituents in a sentence.

Table 1 shows a subset of grammatical relations and the associated role-types. For example, based on the definition of the *noun compound modifier* grammatical relation, the "*dependent*" is any noun that serves to modify the

"*head*" noun. Therefore, any query term that is identified as the "*head*" from a linguistic parser will be annotated as *CoI* and its corresponding dependent will be annotated as *Dc*. A similar process of inference is required to create a generic listing of grammatical relations and their associated role-types for the concept-role mapping process (a complete role-type mapping listing which is adopted in this work is provided in Section 4.1).

**Table 1** Categories of the grammatical relations for the Concept-Role mapping process

|   | Category | Grammatical Relation | Function | Concept Roles | |
|---|---|---|---|---|---|
|   |   |   |   | Head | Dependent |
| 1 | Modifier | Noun Compound Modifier | any *noun* that serves to modify the head *noun* | **CoI** | **Dc** |
| 2 | Argument | Object of Preposition | Head of a *noun* phrase following the *preposition* | **Rc** | CoI |
| 3 | Auxiliary | Conjunction | The relation between the head of an noun phrase, *noun* and its *determiner* | CoI | **Sc** |

To summarize, the concept-role mapping process then entails two consecutive steps: (i) Extraction of grammatical relations between query constituents and determination of "head" and "dependent" role. (ii) Assignment of concept roles depending on the functionality of the constituents linked by the grammatical relations.

**Handling Ambiguous Mapping and Untagged Concepts**

In some instances, two significant issues namely the untagged and ambiguous concept-role mapping may occur depending on the parse output of queries. The first challenge, untagged query concepts, occur when the most generic grammatical relation, undefined (*undef*), is formed due to an inability to define a more precise relation between a pair of terms. A statistical approach that capitalizes on the frequency of a concept to determine untagged concept role types is proposed in this thesis. The frequency of terms can be obtained from Google as it provides the frequency of terms across all indexed documents on the web. This is in line with most existing approaches of identifying key concepts where the assumption is that frequently occurring concepts are key concepts. Thus, three rules are derived to handle untagged concepts:

(i) **Rule 1**: the term that occurs more frequently is the *CoI* and the other is a *Dc*;
(ii) **Rule 2**: if both terms are equally frequent, both terms are tagged as *CoIs*; and lastly
(iii) **Rule 3**: if a term is tagged in one of the defined relations, the untagged term inherits the other terms role.

The application of these rules is demonstrated through the example query "*United States control of insider trading*" in Table 2. A parser derives three relations among the query terms i.e. *undefined* (undef), *noun compound modifier* (nn) and *preposition* (prep). It is observed that the relationship between the terms "*United States*" and "*control*" is undefined (i.e. *undef*). In accordance to Rule 3, the term "*control*" inherits the role-type "*Dc*" from its occurrence in the relation "*prep of*". The term "*United States*" is originally tagged as "U" which means the term could not be mapped to an appropriate role-type. However, upon examination of its frequency in accordance to Rule 1 the term "*United States*" is tagged with the role-type "*Dc*".

**Table 2** Handling Untagged Query Terms

| Relation | Head | Role-Type | Dependent | Role-Type |
|---|---|---|---|---|
| nn | trading | CoI | insider | Dc |
| ***undef*** | ***United_States*** | ***U*** | ***control*** | ***U*** |
| prep_of | control | Dc | trading | CoI |
| prep_of | of | Rc | - | - |

The second challenge, ambiguous concept-role mapping occurs due to the nature of natural language queries in which adjacent and non-adjacent terms may be syntactically/semantically associated. This means a single term can be related to one or more terms within a query, playing a different role in each partnership. To resolve this, the tag of the more significant role is retained. In the case where the contradiction is caused by either *preposition* (*prep*) or *conjunction* (*conj*) relation, the role of the other relation is retained. This is because "*prep*" and "*conj*" serve to concatenate terms and as such it is assumed to be of less significance than other relations. If the contradiction is between "*prep*" and "*conj*", the role from the "*prep*" relation is retained because prepositions connect terms that modify one another while conjunctions simply connect two terms. The retention of the more significant role is illustrated as follows:

(i) **Scenario 1 -** Query: "*efforts to improve United States schooling*" (Table 3).
A parser derives four relations among the query terms i.e. nominal subject (nsubj), auxiliary (aux), noun compound modifier (nn) and direct object (dobj). A contradiction in role assignment is seen with the term "*improve*" where it is labelled as *CoI* and *Dc* across three grammatical relations. In this case, the more significant role i.e. *CoI* is retained.

**Table 3** Handling Ambiguous CR Mapping (Scenario 1)

| Relations | Head | Role-Type | Dependent | Role-Type |
|---|---|---|---|---|
| nsubj | **improve** | ***Dc*** | efforts | CoI |
| aux | **improve** | ***CoI*** | to | Sc |
| nn | schooling | CoI | United_States | Dc |
| dobj | **improve** | ***Dc*** | schooling | CoI |

(ii) **Scenario 2 -** Query: "*Iranian support for Lebanese hostage takers*" (Table 4)
A parser derives two relations among the query concepts i.e. *adjectival modifier (amod)* and *prepositional modifier (prep)*. A contradiction in the role assignment is seen with the concept "*support*". The role i.e. *CoI* from the relation "*amod*" rather than the *Dc* role-type in the the relation "*prep_ for*" is then retained.

**Table 4** Handling Ambiguous CR Mapping (Scenario 2)

| Relations | Head | Role-Type | Dependent | Role-Type |
|---|---|---|---|---|
| amod | **support** | ***CoI*** | Iranian | Dc |
| amod | hostage_takers | CoI | Lebanese | Dc |
| prep_for | **support** | ***Dc*** | hostage_takers | CoI |
| prep_for | for | Sc | - | - |

To illustrate the process of concept-role mapping, we again refer to the query "*coping with overcrowded prisons*" (Figure 2). The linguistic parser returns two grammatical relations between the query terms, i.e. adjectival modifier and prepositional modifier. The relation, adjectival modifier, shows that "*overcrowded*" is an adjective that modifies the noun "*prisons*" while the, prepositional modifier relation, shows that preposition "*with*" modifies the verb "*coping*" and relates it to the noun "*prisons*". Referring to the set of predefined concept-role assignments, we would be able to map each query term to its corresponding role, i.e. "prisons" is a *CoI*, "overcrowded" is a *Dc*, "coping" is a *Dc* and "with" is a *Sc*.

The algorithmic instantiation of the Concept-Role Mapping process is shown in Algorithm 3.1.2. The queries are parsed using a linguistic parser which results in each query term being labelled with POS. Syntactical dependencies are analyzed to derive semantically associated base pairs. Each term within base pairs is annotated with one of the pre-defined role types and anomalies such as untagged and ambiguous are handled according to other relational pairs and term frequency.

**Algorithm 3.1.2** Algorithm for Concept-Role Mapping

---

Requires: Q, the set of queries with isolated NCP
1: FOR all q∈ Q do
2:     parseQuery(q); // POS tagging
3:     *basePairs* = extractTypedDependencies(q); // *basePairs* is a vector of mapped concept pairs
4:     FOR each element *basePairs*[i] REPEAT
5:         Assign role-type based on predefined role-type mapping to each concept in *basePairs*[i]
6:             IF a concept *c* is tagged with the **Untagged** role_type
7:                 IF *c* is tagged in other relational pairs
8:                     Assign the **tagged** role-type to *c*
9:                 END IF
10:            IF *basePairs*[i] concepts have differing frequencies
11:                Assign the **CoI** role-type to the concept with higher frequency
12:                Assign the **Dc** role-type to the concept with lower frequency
13:            ELSE IF *basePairs*[i] concepts have equal frequency
14:                Assign the **CoI** role-type to both concepts
15:            END IF
16:    END FOR
17:    IF a given concept c have **ambiguous** role-types
18:      FOR each element *basePairs*[i] REPEAT
19:          IF c is found
20:              Assign the most **significant** role-type
21:          END IF
22:      END FOR
23:    END IF

---

## 3.2 Expansion Concept Extraction

The very nature of natural language dictates that there are intrinsic relationships between adjacent and non-adjacent terms that reveal semantic notions pertaining to a search goal. However, earlier research fails to fully capitalize on these relationships between query terms which if considered appropriately would improve retrieval performance (discussed in Section 1). Typically, natural language queries are made up of syntactical structures (Chomsky, 1957) which inter-connect terms into grammatically-sound sentences. It is hypothesized that both adjacent and non-adjacent association among query terms can be effectively capitalized from syntactical dependencies within queries. This then translates into meaningful query term pairing for expansion term extraction (discussed in Section 3.1.2)

Extraction of expansion concepts involves two major processes, i.e. identification and selection. Identification of potential expansion concepts is reliant on the dependence model considered in determining base pairs of concepts that will be utilized for extracting co-occurring concepts. In the past, it has been centered on position and proximity of concepts within a query. As discussed in Section 1, query expansion that utilizes bigram pairs of concepts outperforms unigram models. However, in existing approaches, all possible pairs formed by assuming dependence among adjacent and/or non-adjacent concepts are utilized in generating potential expansion concepts. We concur that both adjacent and non-adjacent pairs of concepts should be considered during the query expansion exercise. However, it is our conviction that retrieval performance would be improved further if only adjacent and non-adjacent pairs which are formed as a result of typed dependency labeling (cf. Section 3.1.2) is considered.

The selection of semantically associated base pairs for the term pooling process is further specified by considering only pairs which consists of at least *CoI* or *Dc* role-types as either a head/dependent within the pair. Since *CoIs and Dcs* are terms which represent the search goal, this would ensure that only co-occurring terms that are directly related to the search goal of a query are considered during expansion. Base pairs containing only relational concepts (*Rc*) and structural concepts (*Sc*) are not used in the term pooling process. Table 5 shows the constituents of the query "*coping with overcrowded prisons*" that would be related as a result of typed dependency labeling i.e *overcrowded* ↔ *prisons* and *coping* ↔ p*risons*.

**Table 5** Grammatically related pairs for expansion concept extraction

|   | Grammatical Relation | Concept Roles | | | |
|---|---|---|---|---|---|
|   |   | **Head** | | **Dependent** | |
| 1 | Adjectival Modifier | prisons | CoI | overcrowded | Dc |
| 2 | Prepositional modifier | coping | Dc | prisons | CoI |

These pairs would then be compared against an n-gram corpus to extract co-occurring concepts as potential expansion concepts. Statistically-collocated candidate expansion concepts are extracted through the process of approximate string matching against an n-gram model. An n-gram corpus formed as a result of Internet mining where each sentence within the entire set of web pages is tokenized into word sequences of *n*-length is required for the process of term pooling. The number of times each n-gram occurs in the web is used to determine the relatedness of a set of terms to an original query term. Figure 3 shows a subset of 3-gram tokens along with their observed frequency counts within an n-gram corpus.

| *3-gram Token* | *Frequency* |
|---|---|
| *ceramics company facing* | *145* |
| *ceramics company in* | *181* |
| *ceramics company started* | *137* |
| *ceramics component (* | *76* |
| *ceramics composites ferrites* | *56* |
| *ceramics computer graphics* | *51* |
| *ceramics computer imaging* | *52* |
| *ceramics consist of* | *92* |
| *ceramics collection |* | *59* |
| *ceramics collections ,* | *66* |
| *ceramics collections.* | *60* |
| *ceramics combined with* | *46* |
| *ceramics come from* | *69* |

**Figure 3** A sample of 3-grams tokens from an n-gram corpus

**Table 6** Wildcard Sequences for Approximate String Matching

| 3-gram | 4-gram | 5-gram |
|---|---|---|
| *term1 term2 <\*>* | *<\*> term1 term2 <\*>* | *term1 term2 <\*><\*><\*>* |
| *<\*> term1 term2* | *term1 term2 <\*><\*>* | *<\*> term1 term2 <\*><\*>* |
| *term1 <\*> term2* | *<\*> <\*> term1 term2* | *<\*> term1<\*> term2<\*>* |
|   | *term1 <\*> <\*> term2* | *<\*> term1<\*> <\*>term2* |
|   | *term1<\*> term2 <\*>* | *<\*> <\*> term1 term2 <\*>* |
|   | *<\*>term1<\*> term2* | *<\*> <\*> term1 <\*>term2* |
|   |   | *<\*> <\*> <\*> term1 term2* |
|   |   | *term1 <\*><\*><\*> term2* |
|   |   | *term1 <\*><\*> term2 <\*>* |
|   |   | *term1 <\*> term2 <\*><\*>* |

(Note: <\*> indicate the wildcard entry of related terms)

The set of grammatically-linked base pairs formed through the term dependency modeling process for each query is embedded within a *wildcard sequence*. The wildcard sequence contains both terms within a base pair as well as wildcard placeholders representing the surrounding terms. The patterns of co-occurrence which are considered within wildcard sequences include those in which the related terms appear a priori, posterior-to or amidst the base pairs (*term1* and *term2*), as shown in Table 6. Patterns where the terms in the base pair appear inversed are also considered in the matching process. This is due to the recognition that natural language sentences may be in present, past or future tense form causing the base pairs to appear in varying positions within *n*-gram tokens. These patterns of co-occurrences ensure that all terms within a specified context window of the base pair terms are identified from all the web pages. A search of the n-gram corpus is then done for mining n-gram tokens which match the wildcard sequences.

A large set of unordered n-gram tokens are returned for each base pair of a given query. These tokens of varying lengths are merged to create a collection of related n-grams which are then sorted in descending order according to their frequency of occurrence. The most frequently co-occurring candidate expansion terms are extracted by tokenizing each n-gram into its individual terms (Figure 4). The terms "*overcrowded prisons*" is the base pair whilst "*and*" and "*jails*" are the candidate expansion terms. A limitation of this framework is that NCPs are not detected when processing the terms surrounding the base pairs. This would mean losing out on specific NCPs such as proper names which could bring greater specificity to an expanded query.

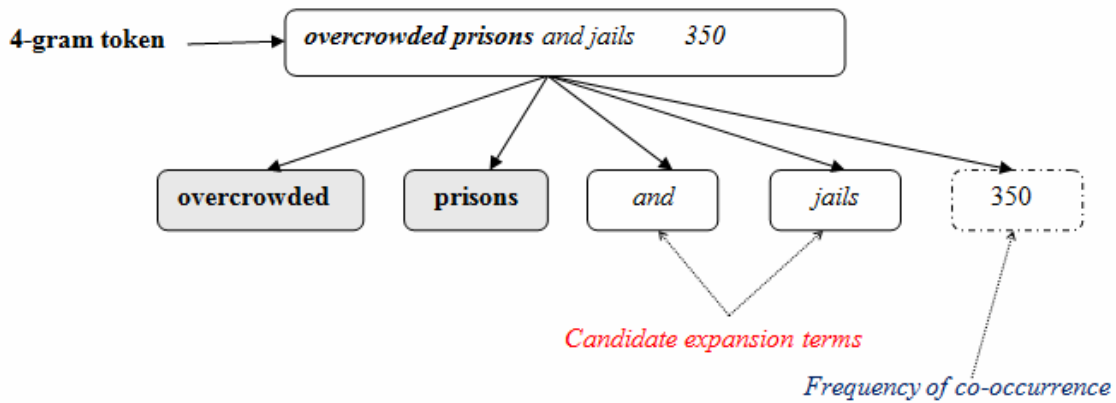

**Figure 4** Tokenized 4-gram token obtained from wildcard sequence matching

The global pool of candidate expansion terms require some processing before specific terms can be selected for use within a final expanded query. A subset of the global pool that is generated for the example query "*coping with overcrowded prisons*" is shown in Figure 5.

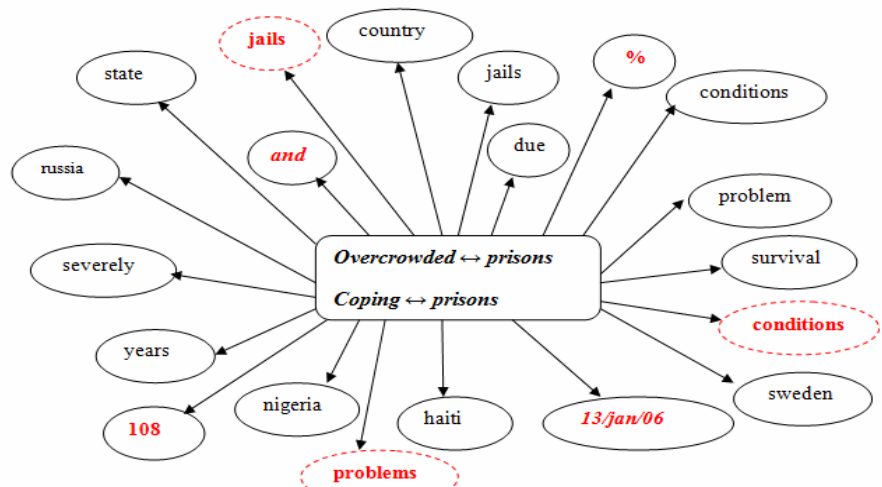

**Figure 5** Subset of pool of candidate terms generated using base pairs of the query "*coping with overcrowded prisons*"

It reveals multiple occurrences of identical terms (e.g. the terms *jails* and *conditions)*, common words/special characters (e.g. and, %), morphological variants (e.g. *problem* and *problems*) and temporal and number expressions (e.g. 108, 13/jan/06) within the generated global pool of candidate terms which may have a negative impact on the performance of the expanded query. Apart from that, it is also necessary to ensure that the candidate expansion terms which match the original query terms are not accepted to avoid redundancy.

Since each query may be associated to several base pairs, the resultant candidate terms may actually overlap. For example, the query "*tobacco company advertising and the young*" has three usable dependencies that are the base pairs: *company* ↔ *advertising, tobacco* ↔ *advertising* and *young* ↔ *advertising*. The base pairs *company* ↔ *advertising* and *tobacco* ↔ *advertising* return several candidate terms among which a subset of them are identical e.g. *billboard, public, marketing*, *online, agencies etc.* In this case, only a single instance of this term is retained within the global pool. Also, morphological variants are commonly found among the extracted terms e.g. *ban, bans, banned.* To deal with morphological variants, the frequencies of individual terms within all web pages are observed and all candidate terms are stemmed to obtain their root form. The individual term frequencies used are unigram frequencies within the n-gram corpus and all the terms are stemmed using Porter's Stemming algorithm (Porter, 1980).

The candidate terms with the same root word are then ranked to determine the most frequently appearing morphological form. The morphological variant with the highest frequency is retained within the global pool. It is acceptable to handle the morphological variants in this manner as the document collections are indexed using the stemmed form of frequently occurring terms. Since all morphological variants would be reduced to their root form, it is sufficient to retain the one form which occurs most often.

Common words, also known as stop words, that exist as structural fillers such as prepositions (e.g. *with, of*), conjunctions (e.g. *and, or*), articles (e.g. *a, the)* and pronouns (e.g. *he, she*) frequently occur but serve no significant purpose in an expanded query. To identify and remove these common words, a standard stop list can be obtained from http://www.ranks.nl/resources/stopwords.html. Special characters consisting of symbols, escape characters, quotation marks, and delimiters should be eliminated. Temporal (e.g. dates and times) and number expressions (money, percentage, and measurements) are also excluded from the global pool of expansion terms. To achieve this, each candidate term is examined to ensure that each character within the term is an alphabet (represented by ASCII codes 65 to 90 for upper case letters and 61 to 122 for lower case). The reasoning behind this decision to exclude such expressions is due to the fact that a document may contain multiple temporal and number expressions relating to a variety of incidents and assessments that may not be central to the intended query intent. If the purpose of such expressions is not properly distinguished, their inclusion could lead to the same effect as using stop words as an expansion term.

The final subset of the global pool for the example query is shown in Figure 6. The number of expansion terms from the generated global pool of statistically-collocated terms which will be included in the final expanded query is experimentally determined and discussed in Section 4.1.

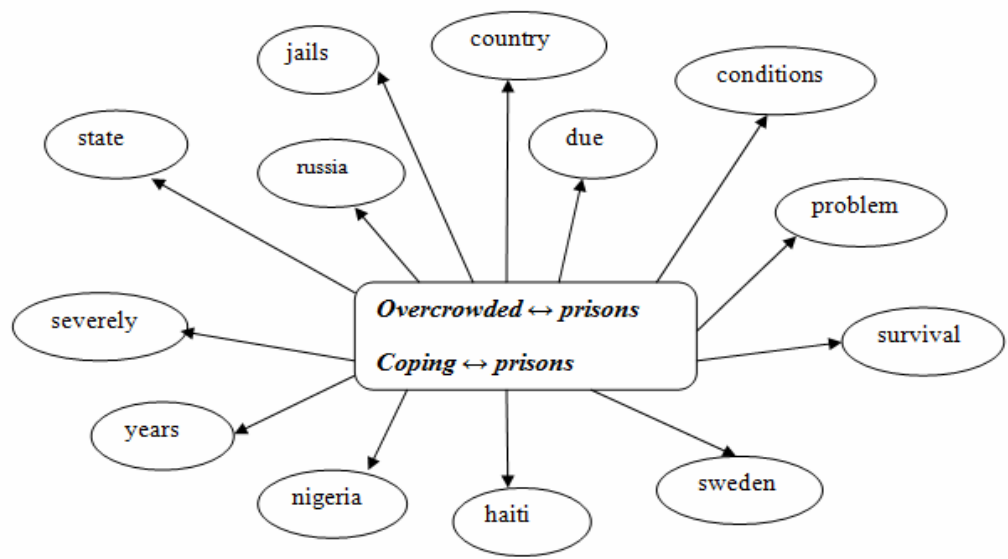

**Figure 6** Final Subset of the pool of candidate terms generated using base pairs of the query "*coping with overcrowded prisons*"

The algorithmic instantiation of the Expansion Concept Extraction Process is shown in Algorithm 3.2

**Algorithm 3.2** Algorithm for Expansion Concept Extraction Process

---
Requires: Q, the set of queries with role-type mapping
1: FOR all q∈ Q do
2:      FOR each element *basePair*[i]
3:           IF a concept is tagged with role-type **CoI**
4:                Store basePair[i] in *significantBasePair* structure
5:           END IF
6:      END FOR
7:      FOR each element *significantBasePair*[i] do
8:           Create *wildCardSeq* set for matching against n-grams
9:           Store in *matchingNgrams* structure the n-grams with identical pattern of co-occurrence
10:          Sort *matchingNgrams* based on frequency of co-occurrence
11:          Tokenize all elements in *matchingNgrams* to extract candidate terms and store in *globalPool* structure
12:     END FOR
13:     FOR each candidate term in *globalPool*
14:          Detect multiple occurrences of identical terms and retain a single entry
15:          Eliminate stop words, special characters, temporal and number expressions
16:          Stem terms to its root form to identify sets of morphological variants
17:          FOR all morphological variants with similar root word
18:               Determine unigram frequency from n-gram corpus
19:          END FOR
20:          Sort variants in descending order of unigram frequency
21:          Store in the *globalUniqueVariantPool* structure the variant with highest unigram frequency
22:     END FOR
23: END FOR
---

All queries which have been processed based on Algorithms 3.1.1 and 3.1.2 are used in this term pooling process. Each query is associated to a set of base pairs. However, only base pairs which contain Concepts-of-Interest (*CoIs*) or Descriptive Concepts (*Dcs*) are considered significant and as such selected for further processing. Wildcard sequences are then created for each base pair. Approximate string matching is conducted for mining n-gram tokens which have identical patterns of co-occurrence. This pool of matched tokens is then ranked in descending order. Based on a cut-off threshold determined experimentally, a select number of tokens are then tokenized to extract the candidate expansion terms. The global pool of candidate expansion terms is then examined to handle several special cases. Firstly, identical terms which appear multiple times are filtered to retain only one instance of the term. Each character within a candidate term is checked to ensure that they only consist of a certain range of ASCII code in order to remove special characters, temporal and number expressions. Stop words are excluded by comparing against an array of standard common words. Morphological variants are identified by stemming all candidate terms according to Porter's Stemming Algorithm (Porter, 1980). Terms with similar root words are collated and their unigram frequency is determined based on unigram frequencies given by an n-gram corpus. The variant with the highest unigram frequency is retained in the global pool of candidate expansion terms.

### 3.3  Concept Weighting

In order to reflect the role of the different query concepts in an expanded query, a weighting mechanism wherein the prior weights assigned are truly reflective of the role of a concept within the expanded query and are in addition to any concept weighting done by retrieval models is required. Each *role-type* is assigned a weight reflective of a prior estimate of their importance. This is based on the assumption that *CoIs* are indicative of the search goal of the query whilst descriptive concepts further specify the search goal. Relational concepts play a smaller role but would still signify the relevance of a document whilst structural concepts are stop words which are common words that do not play a major role in distinguishing document relevance. Although it may be intuitively assumed that *CoIs* and *Dcs* are more important than *Rcs, Scs, and Ecs,* this assumption may not always hold true. The usefulness of a term within a query is not always determined solely by its functionality. Rather, retrieval performance is greatly enhanced by the inclusion of crucial terms where the common theme among this set of terms significantly impacts its retrieval effectiveness (Ogilvie et al., 2009). Thus, a technique that can randomly vary the term weights to optimize performance of a query is formalized through the use of a Genetic Algorithm (GA) formulation. GA is based on biological evolution which simulates survival of the fittest and is widely used as an optimization solution. A GA algorithm mimics a process of natural selection by repeated evolution of a population of candidate solutions in an attempt to find the optimal solution. Each candidate solution (i.e. *individuals*) is tested to determine its relative success (i.e. *fitness*) and the most fit candidate solution is reproduced in the next generation. The search for the optimal solution is achieved through the interaction process with itself (e.g. mutation, crossover) and the environment (e.g. evaluation and selection) (Sanchez et al., 1997).The use of a genetic algorithm based weighting scheme makes it possible to perform optimization by considering a population of chromosomes representing the varying role-types in a given query, rather than single entities like typical continuous optimization techniques (e.g. steepest descent, Newton's method). This would allow a global optimum to be reached and reduces the likelihood of being trapped in a local optimum, thus increasing the robustness of the proposed scheme.

A GA formulation is developed in line with the goal of maximizing mean average precision (MAP) of retrieval by means of predicting the best weights for each role-type for a given query.

The GA optimization consists of the following:

    i. Initialization of the population of chromosomes: a set of chromosomes representing potential solutions consisting of five genes representing each *role-type* (Concepts-of- Interest (*CoI*),(Descriptive Concepts (Dc), Relational Concepts (Rc), Structural Concepts (Sc) and Expansion Concepts(Ec))* is generated. Boundary conditions are specified to limit the range of the weights in accordance to the numeric limitations imposed by the chosen retrieval model. The evolutionary cycle then commences where each potential solution is evaluated in accordance to a predefined fitness function.

    ii. Implementation of a Fitness Function: A fitness function for evaluating each potential solution is defined where the measure of its performance is based on the maximum MAP returned by the retrieval model for a given set of queries. The GA implementation initiates the retrieval process for a set of queries that are equipped with the set of weights obtained from each chromosome in the population. The returned MAP is compared with the retrieval results of each member of the population to establish the highest MAP achieved for the set of queries. The fitness value is used as a basis for the selection process when creating subsequent generations where potential solutions with high fitness value will have high probability of propagating into the next generation.

    iii. Configuration: the fittest individual, with the maximum MAP of the current population is selected through the natural selection process and propagated to the next generation through the process of reproduction. Population is evolved for a fixed number of iterations as a terminating condition. Random perturbations to the population of chromosomes are inflicted through the application of two biological operators, mutation and crossover. The crossover process involves the combination of two potential solutions which high fitness values in an attempt to produce an offspring with even higher fitness. Mutation, on the other hand, involves the random adjustment of a potential solution's genetic structure. The rate of mutation and crossover is experimentally determined as the parameters for optimal GA performance is system-dependent.

    iv. Termination: there are many possible criteria for stopping an algorithm from continued evolution, including number of evolutions, time duration and whether the fitness is continuing to improve. Since the time taken to complete a retrieval process differs for varying dataset sizes, time duration is not chosen as a terminating condition. The continuous improvement of the fitness value is also an inappropriate measure for termination as the changes in achieved MAP is not always consistent where a sudden increase in MAP may be observed after a series of deteriorating MAP outputs. Therefore, the optimal number of evolutions is used as a terminating condition. The maximum evolution parameter is determined experimentally by an observation of when the achievable MAP converges to a reasonably steady-state.

    The implementation of GA for the optimization of weights results in multiple sets of weights representing different role-types for a given query. The query "*coping with overcrowded prisons*" carries the following sample *role-type* weights composition wherein the concept "*prisons (CoI)*" is assigned the weight 0.859, the concept "*with (Sc)*" with is assigned 0.0, the concepts "*overcrowded (Dc)*" and "*coping (Dc)*" are assigned the weight 0.157 whilst all expansion concepts i.e. *state, years, country, conditions, problems etc (Ec)* are given equivalent weights of 0.064. Although a pattern of weighting is revealed in this sample where the weights of the role-types descend in the following order, *CoI > Ec > Sc,* it is not possible to generalize that all queries will be having the same patterns of weight.

The algorithmic instantiation of the concept weighting process is shown in Algorithm 3.3.

**Algorithm 3.3** Algorithm for Expansion Concept Extraction Process

---

Requires: Q, the set of queries with role-type mapping
1: FOR all q∈Q do
2:        Initialize population with 5 role-types and bounded weights
3:        Evaluate population based on highest Mean Average Precision (MAP) achieved
4:        Determine optimal crossover rate, mutation rate and maximum evolution
5:        WHILE (! maximum evolutions)
6:              Select potential solutions with high fitness value for reproduction
7:                  Implement crossover and mutation
8:              Initiate retrieval process for current population
9:              Evaluate population based on highest MAP achieved
10:       END WHILE
11: END FOR

---

The expanded query is examined on its effectiveness by application of role-type weights on all query terms. Through a GA implementation, an initial population of chromosomes representing the possible role-type weights (e.g. *CoI, Dc, Rc, Sc, EcStat*) are generated. The final-expanded queries are annotated with term weights reflective of the term role-type within a query derived from the GA optimization system.

## 4 Experimental Validation

In this section, we present the experimental setup and results for each process defined in the framework. Section 4.1 describes the experimental setup for the evaluation of our framework and in Section 4.2 we present the retrieval results. We evaluate our model on the TREC ad hoc test collection topics (Tipster Disk 1&2 consisting of documents from the Wall Street Journal, Associated Press, Federal Register, Department of Energy and Computer Select as well as Gov2). We examine the title field of 200 queries in Trec 1, 3, 8 and Terabyte 2005 of the ad hoc test topics (Table 7).

**Table 7** Summary of Trec Collections & Topics

|  | # Docs | Topics |
|---|---|---|
| WSJ90_92 | 74,520 | 51-100 |
| AP88-90 | 242,918 | 51-100 |
| SJM1991 | 90,257 | 51-100 |
| WSJ87_92 | 173,252 | 151-200 |
| AP88_89 | 164,597 | 151-200 |
| Disk4-5 (MinusCR) | 489,164 | 401-450 |
| Gov2 | 25,114,919 | 751-800 |

The Stanford NLP parser (*available at http://nlp.stanford.edu/software/*) was utilized for all linguistic processing tasks e.g. part-of-speech tagging and derivation of grammatical relations. Queries have to be appropriately formatted to ensure accurate parse trees are generated from the parser. In the case of the Stanford NLP parser, the components of the NcPs were joined with the underscore to represent the phrase as one unit. Proper names are capitalized, front slashes (/) were substituted with the word "or", hyphenated and bracketed words were retained and double quotes were dropped and replaced by underscores to encase the concepts. Acronyms were not only isolated but also resolved into its full form.

Indexing and retrieval were performed utilizing the Lemur-Indri Toolkit (*available at http://www.lemurproject.org/indri/*) which implements a language model. The optimal Dirichlet smoothing parameter was determined experimentally where the highest retrieval performance was obtained when its value was set at $\mu$=1500. Stop words were retained and all terms were stemmed. The Indri model defines belief operators (described in http://www.lemurproject.org/lemur/IndriQueryLanguage.php). Belief operators allow for beliefs (scores) about terms, phrases, etc. With the #weight operator, varying weights can be assigned to certain expressions (terms, phrases, etc). The impact each expression within the query has on the final score is controlled through the computa-tion of weighted geometric average. Further, the ordered window operator, #N( ), is used to encapsulate NCP terms found within the original query terms, where the terms must appear ordered, with at most N-1 terms between each term (Strohman et al., 2005). Such multi-word expressions are not considered from within the expansion terms and it is acknowledged as a limitation of the work. The final expanded query includes the weights ($W_t$) of the five role-types defined in Section 3.3 where $W_t$ may be $W_{CoI}$, $W_{Dc}$, $W_{Rc}$, $W_{Sc}$, or $W_{Ec}$ as shown in the query format below:

>#weight (
>
>>$W_t$ term1
>>$W_t$ term2
>>$W_t$ term3
>>$W_t$ term4
>>$W_t$ #1(term5 term 6)
>
>)

A paired-samples t-test is conducted to analyze the significance of the gained retrieval performance of all experimental variations.

## 4.1 Setup

The queries were compared against the set of NcPs (i.e. phrasal verbs, collocations, idioms, proper names and acronyms) contained in an external knowledge base (Wordnet ontology) to isolate non-compositional phrases. Since the knowledge-base may not comprehensively cover all possible NcPs (especially proper names), the queries were manually parsed to identify any remaining undetected NcPs by referencing lists of NcPs on the web. Query constituents are annotated in accordance to the concept-role mapping process defined in Section 3.1.2. For this purpose, the functionality of the query constituents linked by the grammatical relations was deduced based on the description of the relation that holds between a head and a dependent, as defined in defined in the Stanford Typed Dependency Scheme (Marneffe & Manning, 2008) . These typed dependencies capture semantic associations among adjacent and non-adjacent query constituents and are used to determine role-types of the query terms during the concept-role mapping process. A brief description of the grammatical dependencies (Marneffe & Manning, 2008) within three categories of relations (arguments, auxiliaries, modifiers) that were used in this research and the corresponding role-type mappings is provided in Table 8.

**Table 8** Listing of Grammatical Relations & Role-Type Mapping

| Category | Grammatical Relations | Description | Head | Dependent |
|---|---|---|---|---|
| **Arguments** | coordination | A coordination is the relation between an element of a conjunct and the coordinating conjunction word of the conjunct. | CoI | Rc |
| | adjectival complement | An adjectval complement of a verb is an adjectival phrase which functions as the complement. | Dc | CoI |
| | clausal complement | A clausal complement of a verb or adjective is a dependent clause with an internal subject which functions like an object of the verb, or adjective. | Dc | CoI |
| | open clausal complement | An open clausal complement of a VP or an ADJP is a clausal complement without its own subject, whose reference is determined by an external subject. | Dc | CoI |
| | complementizer | A complementizer of a clausal complement is the word introducing it. It will be the subordinating conjunction "that" or "whether". | CoI | Rc |
| | direct object | The direct object of a VP is the noun phrase which is the (accusative) object of the verb. | Dc | CoI |
| | indirect object | The indirect object of a VP is the noun phrase which is the (dative) object of the verb. | Dc | CoI |
| | object of a | The object of a preposition is the head of a noun | Rc | CoI |
| | preposition | phrase following the preposition, or the adverbs | | |
| | marker | A marker of an adverbial clausal complement is the word introducing it. | CoI | Rc |
| | relative | A relative of a relative clause is the head word of the WH-phrase introducing it. | CoI | Rc |
| | nominal subject | A nominal subject is a noun phrase which is the syntactic subject of a clause. | Dc | CoI |
| | Passive nominal subject | A passive nominal subject is a noun phrase which is the syntactic subject of a passive clause. | Dc | CoI |
| | clausal subject | A clausal subject is a clausal syntactic subject of a clause, i.e., the subject is itself a clause. | Dc | CoI |
| | clausal passive subject | A clausal passive subject is a clausal syntactic subject of a passive clause. | Dc | CoI |
| | expletive | This relation captures an existential "there". | Rc | Rc |
| | prepositional complement | This is used when the complement of a preposition is a clause or prepositional phrase (or occasionally, an adverbial phrase). | Rc | Rc |
| | preconjunct | A preconjunct is the relation between the head of an NP and a word that appears at the beginning\bracketing a conjunction (and puts emphasis on it), such as "either", "both", "neither"). | CoI | Rc |
| **Modifiers** | Noun compound modifier | any noun that serves to modify the head noun | CoI | Dc |
| | adjectival modifier | An adjectival modifier of an NP is any adjectival phrase that serves to modify the meaning of the NP. | CoI | Dc |
| | prepositional modifier | A prepositional modifier of a verb, adjective, or noun is any prepositional phrase that serves to modify the meaning of the verb, adjective, noun, or even another prepositon. | Dc | CoI |
| | abbreviation modifier | An abbreviation modifier of an NP is a parenthesized NP that serves to abbreviate the NP | CoI | CoI |
| | appositional modifier | An appositional modifier of an NP is an NP immediately to the right of the first NP that serves to define or modify that NP. It includes parenthesized examples. | CoI | CoI |
| | adverbial clause modifier | An adverbial clause modifier of a VP or S is a clause modifying the verb (temporal clause, consequence, conditional clause, etc.). | Dc | CoI |
| | purpose clause modifier | A purpose clause modifier of a VP is a clause headed by "(in order) to" specifying a purpose. | Dc | CoI |
| | numeric modifier | A numeric modifier of a noun is any number phrase that serves to modify the meaning of the noun. | CoI | Dc |
| | element of compound number | An element of compound number is a part of a number phrase or currency amount. | CoI | Dc |
| | possession modifier | The possession modifier relation holds between the head of an NP and its possessive determiner, or a genitive 's complement. | CoI | CoI |
| | phrasal verb particle | The phrasal verb particle relation identifies a phrasal verb, and holds between the verb and its particle. | Dc | CoI |
| | parataxis | The parataxis relation (from Greek for "place side by side") is a relation between the main verb of a clause and other sentential elements, such as a sentential parenthetical, or a clause after a ":" or a ";". | CoI | Rc |
| | punctuation | his is used for any piece of punctuation in a clause, if punctuation is being retained in the typed dependencies. | CoI | Sc |
| | referent | A referent of the head of an NP is the relative word introducing the relative clause modifying the NP. | CoI | Rc |
| | controlling subject | A controlling subject is the relation between the head of a open clausal complement (xcomp) and the external subject of that clause. | Dc | CoI |
| **Auxiliaries** | auxilliary | An auxiliary of a clause is a non-main verb of the clause, e.g. modal auxiliary, "be" and "have" in a composed tense. | CoI | Rc |
| | passive auxiliary | A passive auxiliary of a clause is a non-main verb of the clause which contains the passive information. | CoI | Sc |
| | copula | A copula is the relation between the complement of a copular verb and the copular verb. | CoI | Rc |
| | agent | An agent is the complement of a passive verb which is introduced by the preposition "by" and does the action. | Rc | CoI |
| | attributive | An attributive is a complement of a copular verb such as "to be", "to seem", "to appear". | Rc | Rc |
| | conjunction | A conjunct is the relation between two elements connected by a coordinating conjunction, such as "and", "or", etc. | CoI | CoI |
| | determiner | A determiner is the relation between the head of an NP and its determiner. | CoI | Sc |
| | predeterminer | A predeterminer is the relation between the head of an NP and a word that precedes and modifies the meaning of the NP determiner. | CoI | Rc |

    To capture both adjacent and non-adjacent semantic relations, we capitalize on the typed dependencies obtained through the NLP parser. The parser returns semantically associated "heads" and "dependents" based on the modified Collins head rules. We utilize Google n-grams *(available at http://www.ldc.upenn.edu/Catalog/CatalogEntry.jsp?catalogId=LDC2006T13)* as a document corpus for extraction of expansion concepts with a co-occurrence window of maximum size 5. Wildcard sequences were designed to use the wildcard matching property within the Get1t-03 software to retrieve all possible co-occurring matches with the base pairs. Three sets of all possible wildcard permutations were created based on the sequences presented in Section 3.2. The search of the n-gram corpus for each base pair through the process of approximate string matching returns a subset of related n-gram tokens. The number of matched entries returned from approximate string matching of the wildcard sequences varies depending on the degree of specificity of the base pairs. For example, the base pair "*information technology*" returns more than two hundred thousand entries as compared to the base pair "*overcrowded prisons*" which returns less than one hundred matching n-gram tokens. Each entry in the retrieved subset of matched n-grams was then tokenized to gather all surrounding terms. Naturally, a large global pool of related terms was formed as a result. Identical terms, common terms, special characters, morphological variants, temporal and number expressions were then removed. For the multiple morphological variations of concepts generated, the frequency of all concepts which share the same root word is summed and these concepts are sorted in descending order for selection of *Top n* concepts for expansion. The Porter's Stemming algorithm (Porter, 1980) was used to determine the root word of each expansion concept. Frequencies of concepts are obtained from Google N-grams. In the case of missing concepts due to the high threshold of cut-off in the Google n-gram corpus, we obtain the required

frequency counts directly from the Google search engine. Due to the differing sources of frequency counts, scores are normalized by performing a proportional normalization.

We select the *Top 5* concepts for expansion from the local pool of co-occurring concepts which were collated from all matches to base pairs for a given query and sorted based on the concept frequency within Google N-grams. The optimal number of candidate expansion terms ($t_c$) to be included in a final expanded query was determined by observing retrieval performance over several increments of total expansion concepts where $t_c$ = 5, 10, 20, 30, 40, 50 and 100 in the expansion process. Small increases in MAP were obtained as the value of $t_c$ was increased from 5 to 30 terms, with deterioration in retrieval performance shown when $t_c$ was set at 40, 50 and 100. The increase in MAP observed as terms were added on from 5 to 30 candidate terms was not statistically significant on a set of 50 queries (Trec3 queries). Therefore, the optimal number of candidate terms was set to 5 for the experiments conducted.

Role-type weights are predicted and optimized through the use of Genetic Algorithm (GA). Role weights are generated for 4 out 5 role types identified e.g. CoI, Dc, Rc and Ec with a boundary condition set to restrict weights within a 0.000 to 1.000 weight range. The Sc role type is enforced with a weight of 0.000 in order to eliminate emphasis on common concepts as described in Section 3.1.2. The parameters and optimal rates utilized in our GA implementation were determined through a tuning phase by varying the parameters on 50 query test cases. The GA implementation for maximization of average precision includes the mutation and crossover genetic operators which are set at a rate of 10 and 1000 respectively. A population 80 chromosomes representing the role-type weights is evolved over 100 iterations and serves as a termination condition. The fitness value of each chromosome is boosted based on the overall MAP generated e.g. Fitness Value is boosted if MAP is above 0.1, above 0.2, above 0.3, above 0.4 and above 0.5 with the highest boost given to the highest MAP (above 0.5) across all iterations. The fitness value is used as a basis in the fitness proportionate selection method where a roulette-wheel selection technique is employed for choosing the chromosomes with high fitness value for reproduction and propagation to subsequent evolutions. The returned MAP from the retrieval model is compared with the retrieval results of each member of the population to establish the highest MAP achieved for the set of queries.

## 4.2 Retrieval Results

In order to assess the performance of the proposed framework, it is recognized that the experimental results obtained must be compared against recent research efforts in the field. However, the inconsistencies in the experimental setup of current research prevent a fair comparison. It is noted that in existing research efforts the accepted de-facto standard for comparison are the unigram language model (LM) (Ponte & Croft, 1998) and relevance model (RM) (Lavrenko & Croft, 2001). An analysis of recent works in the field by Carpineto & Romano (2012) revealed large differences on the baseline MAP results reported even on similar test sets. The reason for these inconsistencies is due to a variety of factors including variations in optimization parameters selected in the running of specific retrieval engines (e.g. LEMUR, UMASS, SMART, LUCENE, etc) such as smoothing parameters, ranking methods, stop word elimination, stemming etc at the point of indexing and retrieval. As such, even though the test collections are identical in terms of its queries examined and document set size, the achievable baseline results are often different. Since this is the case, the performance of the proposed approach cannot be assessed by comparing against the absolute MAPs achieved in existing research. Therefore, in line with the practice of current research work, we examine the performance of our framework (LSQE) by comparing against the state-of-the-art unigram language model (LM) (Ponte & Croft, 1998), relevance model (RM) (Lavrenko & Croft, 2001). The language model implementation represents the baseline performance of the original query without any form of manipulation. On the other hand, the implementation of the relevance model signifies the baseline performance of the state-of-the-art pseudo relevance feedback technique which incorporates the top n frequently occurring terms which are obtained from the top k documents returned for an initial query. A comparison is also made with the sequential processing based query expansion (SPQE) technique. The SPQE framework differs from LSQE, firstly in the way the base pairs are formed for the purpose of expansion concept extraction. In LSQE, we rely on grammatical dependencies to form head-dependent pairs (as described in Section 3.1.2 and 4.1) whilst in SPQE, base pairs are formed based on the notion of term dependencies, specifically sequential dependence. Emulating sequential dependence (Metzler & Croft, 2005) where dependence is assumed to exist between adjacent query terms, sequentially dependent base pairs were then formed for the purpose of expansion concept extraction in SPQE. Secondly, the final expanded query consists of original and expansion concepts weighted via GA without varying weights for each role of query constituents as done in LSQE.

We present two variations in query selection for the purpose of evaluation. In Query Selection 1 (QSE1), we exclude 1 word queries, queries with zero relevance judgments as well as un-expandable queries. Our proposed approach does not cater for one-word queries as we rely on grammatical pairing of concepts during the process of expansion concept extraction. Un-expandable queries are queries with no expansion concepts remaining after the process of filtering during the LSQE processes. We exclude such queries in all experimental variations to allow for fair performance comparison. We evaluate retrieval performance by comparing the improvements in Mean Average Precision (MAP) obtained over LM, RM and SPQE. The MAP results of this query set (QSE1) are presented in Table 9 and the relative improvements are shown in Figure 7. A paired-samples t-test was conducted to evaluate statistical significance of the improvements in retrieval performance over LM, RM and SPQE (indicated in Table 9).

**Table 9** Retrieval performance of LM, RM, SPQE and LSQE (QSE1). The superscripts ¥, † and α indicate statistically significant improvements ($p < 0.05$) over LM, RM and SPQE, respectively.

| | | | QSE1 | | | |
|---|---|---|---|---|---|---|
| Query No | Dataset | Total Queries | LM (MAP) | RM (MAP) | SPQE (MAP) | LSQE (MAP) |
| 51-100 | WSJ90-92 | 40 | 0.1636 | 0.1696 | 0.1734$^{¥}$ | 0.2011$^{¥†α}$ |
| | AP88-90 | 41 | 0.1615 | 0.1653 | 0.1687$^{+}$ | 0.2037$^{¥†α}$ |
| | SJM1991 | 40 | 0.1304 | 0.1421 | 0.1334$^{¥}$ | 0.1764$^{¥†α}$ |
| 151-200 | WSJ87-92 | 47 | 0.239 | 0.2981 | 0.2543 | 0.3097$^{¥}$ |
| | AP88-89 | 47 | 0.2644 | 0.3142 | 0.2834 | 0.3285$^{¥}$ |
| 401-450 | Disk 4-5(-CR) | 40 | 0.1851 | 0.2115 | 0.2005$^{¥}$ | 0.2142$^{¥†α}$ |
| 751-800 | GOV2 | 46 | 0.302 | 0.3145 | 0.3349$^{¥}$ | 0.3389$^{¥}$ |

Both SPQE and LSQE show increase in MAP over LM for each dataset examined. SPQE displayed relative improvements over LM in the range of 2.4 – 10.9% while LSQE showed greater increase in MAP over LM within a range of 12.3 -35.3% across the various datasets. The improvements over LM shown are all statistically significant for LSQE (LSQE_LM) whilst significance is reported for SPQE (SPQE_LM) on four datasets. These improvements over LM support our belief that the use of bigram base pairs in the process of expansion concept selection can improve retrieval performance. PQE outperforms RM within a range of 2.1-6.5%, however, it deteriorates over several datasets between 5.2-14.7%. This poor performance is likely due to the use of sequentially formed base-pairs which possibly resulted in the extraction of irrelevant expansion concepts. Unlike SPQE, LSQE showed 1.3-24% improvements over the relevance model (LSQE_RM) in all datasets with statistical significance present in four of the seven experimental variations. LSQE also demonstrates relative improvements between 1.2 to 32.3% over SPQE (LSQE_SPQE) across all datasets with statistical significance shown on several of the datasets. This range of increase in MAP obtained over SPQE validates our assertion that linguistic-based query constituent encoding is crucial towards improved retrieval. We attribute the better performance of LSQE over SPQE to the use of grammatical dependencies for base pairs formation and the appropriate emphasis on the specific role of the expanded query constituents. Further, the results show that retrieval performance improves with LSQE for both small and large collections.

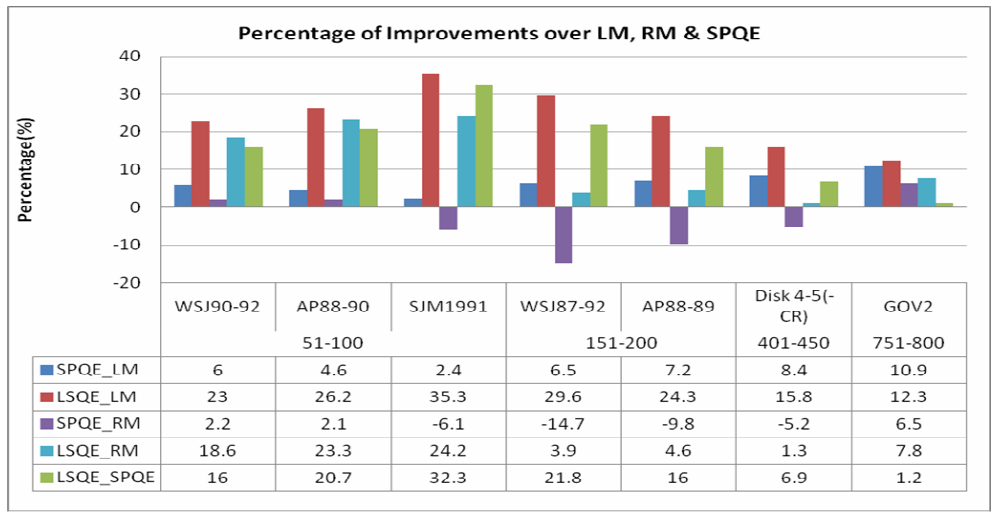

**Figure 7** Relative Improvements over LM, RM & SPQE

We observe improvement in MAP over all datasets. However, we find that the overall MAP achieved is low. Further scrutiny revealed that there are several underperforming queries causing the overall MAP on each dataset to remain low even with QE. We define under-performing queries as those which have low mean average precision (MAP) at baseline and the MAP remains below 0.1 even with LSQE. Thus, in Query Selection 2 (QSE2), shown in Table 10, we also exclude the queries which under-perform to have a clearer picture of the improvements through LSQE and the impact of the under-performing queries on the overall retrieval results. A range of 8-124% increase in overall MAP is observed across the datasets when the underperforming queries were excluded. The marked improvement in the QSE2 results iterates that LSQE fairs well for some portion of the queries across the datasets.

**Table 10** Retrieval performance of LSQE and LM (QSE2).

| Query No | Dataset | Total Queries | QSE2 | |
|---|---|---|---|---|
| | | | LM (MAP) | LSQE (MAP) |
| 51-100 | WSJ90-92 | 21 | 0.2898 | 0.3555 |
| | AP88-90 | 20 | 0.3032 | 0.3882 |
| | SJM1991 | 16 | 0.2948 | 0.3949 |
| 151-200 | WSJ87-92 | 37 | 0.2938 | 0.3789 |
| | AP88-89 | 36 | 0.3359 | 0.4157 |
| 401-450 | Disk 4-5(-CR) | 27 | 0.2578 | 0.299 |
| 751-800 | GOV2 | 42 | 0.3278 | 0.3655 |

Retrieval effectiveness improves for 40-91% of the queries depending on the dataset, which is to say that 65% of the queries have better MAP with LSQE (shown in Figure 8).

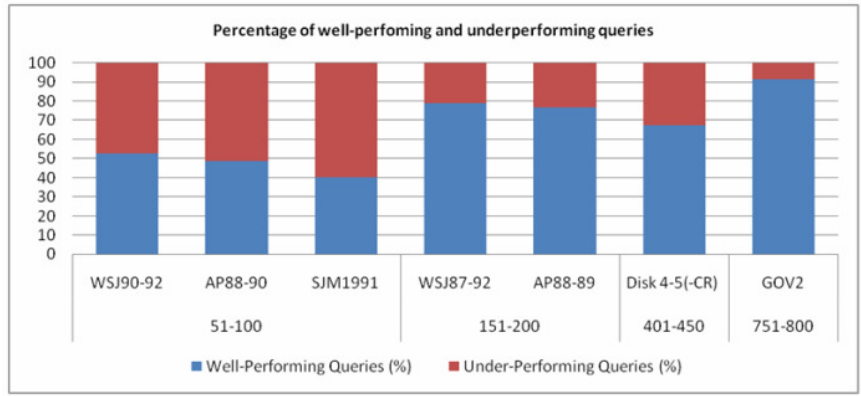

**Figure 8** Percentage of well-performing and under-performing queries

Further, we sorted the experimental dataset based on the query lengths to observe the performance of three groups of varying query lengths i.e. short, medium and long. Short queries consist of no more than 2 words whilst medium queries consist of 3-4 words and all other queries are classified as long queries. Among the remaining 35% of underperforming queries, on average, 2 were short queries, 6 were medium and another 6 were long queries (shown in Table 11). However, among these queries, only 6 queries actually deteriorated in retrieval performance within the LSQE framework. This is indicative of the robustness of the LSQE framework. However, the increase observed in the rest of the queries was minimal whereby the MAP remained below 0.1. We note that even though a similar number of relevant documents were retrieved when these queries perform well within other datasets, for these under-performing queries, the P@N values are typically poor within the first 40-100 retrieved documents, thus resulting in lower MAP.

**Table 11** Number of underperforming queries; shown in brackets, the number of queries that deteriorate with QE

| Query No | Dataset | Number of Queries | Short | Medium | Long |
|---|---|---|---|---|---|
| 51-100 | WSJ90-92 | 19 | 1 | 9 | 9 (1) |
|  | AP88-90 | 21 | 2 (1) | 11 (1) | 8 |
|  | SJM1991 | 24 | 4 (1) | 11 | 9 |
| 151-200 | WSJ87-92 | 10 | 0 | 3 | 7 (1) |
|  | AP88-89 | 11 | 0 | 2 | 9 |
| 401-450 | Disk 4-5(-CR) | 13 | 5 | 8 | 0 |
| 751-800 | GOV2 | 4 | 0 | 3 | 1(1) |

Even though several medium and long queries were underperforming, a large portion of these varying query lengths perform well with LSQE. We tabulate the results of QSE2, sorted by the query length (Table 12). On average, short queries have an overall MAP of 0.4082, which is a 19.9% increase over baseline; medium queries have an average MAP of 0.3703, which is a 23.5% increase over baseline; and 0.3397 for long queries, which is a 32.3% increase over baseline.

## 5 Conclusion

In this paper, we have proposed a framework in which key concepts of queries are identified and encoded through the process of linguistic analysis. In line with our postulation that grammatical pairing of concepts for the process of expansion concept extraction and role-type based weighting are essential in query expansion, we have demonstrated improvements in retrieval performance with our proposed approach. Even though the expansion term extraction strategy can lead to "noisy" concepts, experimentally the gain in terms of MAP with respect to the state-of-the-art techniques is significant. Specifically, we see that varying lengths of queries, inclusive of long queries, show good retrieval performance. We believe that further retrieval performance gains may be observed if the semantic relatedness of expansion concepts to the original query constituents is assessed in the process of expansion concept selection to ensure only closely related concepts are included in the expanded query. In our future work, we will further investigate gains in retrieval performance by incorporating local selection of expansion concepts and measures of semantic relatedness in our linguistic-based query expansion framework.

**Table 12** Performance of varying query lengths; the number of queries is specified in brackets.

|  | Query No | Dataset | Number of Queries | QSE2 | | |
|---|---|---|---|---|---|---|
|  |  |  |  | Short | Medium | Long |
| Basic LM | 51-100 | WSJ90-92 | 21 | 0.3029(10) | 0.2484(6) | 0.3134(5) |
|  |  | AP88-90 | 20 | 0.3713(9) | 0.2305(4) | 0.2572(6) |
|  |  | SJM1991 | 16 | 0.3699(7) | 0.2296(3) | 0.2397(6) |
|  | 151-200 | WSJ87-92 | 37 | 0.2993(4) | 0.374(5) | 0.2787(28) |
|  |  | AP88-89 | 36 | 0.4138(4) | 0.4842(6) | 0.2896(26) |
|  | 401-450 | Disk4-5(-CR) | 27 | 0.3084(15) | 0.1946(12) | n/a |
|  | 751-800 | GOV2 | 42 | 0.3177(11) | 0.337(30) | 0.1619(1) |
| LSQE | 51-100 | WSJ90-92 | 21 | 0.3418(10) | 0.3072(6) | 0.4411(5) |
|  |  | AP88-90 | 20 | 0.4401(9) | 0.3053(4) | 0.3688(6) |
|  |  | SJM1991 | 16 | 0.4765(7) | 0.3393(3) | 0.3274(6) |
|  | 151-200 | WSJ87-92 | 37 | 0.3814(4) | 0.4989(5) | 0.3571(28) |
|  |  | AP88-89 | 36 | 0.4862(4) | 0.5293(6) | 0.3786(26) |
|  | 401-450 | Disk4-5(-CR) | 27 | 0.3378(15) | 0.2504(12) | n/a |
|  | 751-800 | GOV2 | 42 | 0.3939 (11) | 0.3618 (30) | 0.1651(1) |

# References

Abdelali, A., Cowie, J. and Soliman, H. S. (2007). Improving query precision using semantic expansion. Information Processing & Management, 43(3): 705-716.

Aula, A. (2003). Query formulation in web information search. Proc. IADIS Conference WWW/Internet, 403-410.

Bai, J. et al. (2005). Query expansion using term relationships in language models for information retrieval Word sense disambiguation in queries. Proc. ACM SIGIR, 688 - 695

Bendersky, M. and Croft, B. W. (2008). Discovering key concepts in verbose queries. Proc. ACM SIGIR, 491-498.

Bhogal, J., Macfarlane, A. and Smith, P. (2007). A review of ontology based query expansion. Information Processing & Management, 43(4): 866-886.

Chomsky, N. (1957). Syntactic Structures. Mouton de Gruyter.
Cao, G., Nie, J.-Y., Jing, B. (2005). Integrating word relationships into language models. Proc. ACM SIGIR, 298 – 305.

Carpineto, C., and Romano, G. (2012). A Survey of Automatic Query Expansion in Information Retrieval. ACM Computing Surveys, 44(1), 1-50.

Covington, M. A. (2001). A fundamental algorithm for dependency parsing. Proc. Annual ACM Southeast Conference, 95-102.

Di Buccio, E., Melucci, M. and Moro, F. (2014). Detecting verbose queries and improving information retrieval. Information Processing Management. 50(2), 342-360.

Eibe, F., Paynter, G. W., Witten, I. H., Gutwin, C. and Nevill-Manning, C. G. (1999). Domain-specific keyphrase extraction. Proc. IJCAI, 668-673.

Gao, J., Nie, J.-Y., Wu, G. and Cao, G. (2004). Dependence language model for information retrieval. Proc. ACM SIGIR, 170 – 177.

Grineva, M., Grinev, M. and Lizorkin, D. (2009). Extracting key terms from noisy and multitheme documents. Proc. World Wide Web, 661-670.

Hasan, K. S., and Ng, V. (2014). Automatic keyphrase extraction: A survey of the state of the art. Proc. ACL, 1262-1273

Huston, S. and Croft, B. W. (2014). A Comparison of Retrieval Models using Term Dependencies. Proc. ACM CIKM, 111-120

Jing, B., Song, D., Bruza, P., Nie, J.-Y. and Cao, G. (2005). Query expansion using term relationships in language models for information retrieval. Proc. ACM CIKM, 688 – 695.

Kim, S.-B., Seo, H.-C. and Rim, H.-C. (2004). Information retrieval using word senses: root sense tagging approach. Proc. ACM SIGIR, 258 – 265.

Lau, E. P., and Goh, D. H.-L. (2006). In search of query patterns: A case study of a university OPAC. Information Processing & Management, 42(5), 1316-1329.

Lavrenko, V. and Croft, B. W. (2001). Relevance-Based Language Models. Proc. ACM SIGIR, 120-127.

Lioma, C. and Ounis, I. (2008). A syntactically-based query reformulation technique for information retrieval. Information Processing & Management , 44(1), 143-162.

Liu, S., Liu, F., Yu, C. and Meng, W. (2004). An effective approach to document retrieval via utilizing WordNet and recognizing phrases. Proc. ACM SIGIR, 266 – 272.

Liu, S., Yu, C. and Meng, W. (2005). Word sense disambiguation in queries. Proc. ACM CIKM, 525 – 532.

Marneffe, M.-C., Maccartney, B. and Manning, C. D. (2006). Generating Typed Dependency Parses from Phrase Structure Parses. Proc. LREC.

Marneffe, M. -C. and Manning, C. D. (2008). Stanford typed dependencies manual. Stanford University, Techn cal report.